\documentclass[useAMS,usenatbib]{mn2e}
\usepackage{graphicx}
\newcommand{\rcirc}{R_{\rm{circ}}}
\newcommand{\eqref}[1]{(\ref{#1})}
\newcommand{\rschwarz}{R_\mathrm{s}}
\usepackage{times}
\usepackage{hyperref}
\usepackage{booktabs}
\hypersetup{colorlinks=true,citecolor=blue,linkcolor=black,filecolor=black,runcolor=black}

\title[Accretion disc response to black hole mass loss]{Response of a circumbinary accretion disc to black hole mass loss}
\author[Rosotti, Lodato \& Price]{Giovanni P. Rosotti\thanks{E-mail:
rosotti@usm.lmu.de}$^{1,2,3,4}$, Giuseppe Lodato$^{2}$ and Daniel J. Price$^{5}$\\
$^{1}$Max-Planck-Institut f\"ur extraterrestrische Physik, Giessenbachstra\ss{}e, D-85748 Garching, Germany\\
$^{2}$Universit\`a degli Studi di Milano, Dipartimento di Fisica, via Celoria 16, I-20133 Milano, Italy\\
$^{3}$ Excellence Cluster Universe, Boltzmannstr. 2, D-85748 Garching, Germany\\
$^{4}$ Universitats-Sternwarte M\"unchen, Scheinerstra\ss{}e 1, D-81679
M\"unchen, Germany \\
$^{5}$Monash Centre for Astrophysics (MoCA) and School of Mathematical Sciences, Monash University, Vic 3800, Australia}
\begin{document}

\date{Accepted 2012 June 11. Received 2012 June 05; in original form 2012 April 02}

\pagerange{\pageref{firstpage}--\pageref{lastpage}} \pubyear{2011}

\maketitle

\label{firstpage}

\begin{abstract}
We investigate the evolution of the surface density of a circumbinary accretion disc after the mass loss induced by the merger of two supermassive black holes. We first introduce an analytical model, under the assumption of a disc composed of test particles, to derive the surface density evolution of the disc following the mass loss. The model predicts the formation of sharp density peaks in the disc; the model also allows us to compute the typical timescale for the formation of these peaks. To test and validate the model, we run numerical simulations of the process using the Smoothed Particle Hydrodynamics (SPH) code PHANTOM, taking fluid effects into account. We find good agreement in the shape and position of the peaks between the model and the simulations. In a fluid disc, however, the epicyclic oscillations induced by the mass loss can dissipate, and only some of the predicted peaks form in the simulation. To quantify how fast this dissipation proceeds, we introduce an appropriate parameter, and we show that it is effective in explaining the differences between the analytical, collisionless model and a real fluid disc.
\end{abstract}

\begin{keywords}
accretion, accretion discs  -- black hole physics -- hydrodynamics.
\end{keywords}

\section{Introduction}

Supermassive black holes (SMBHs) are hosted in the nuclei of most galaxies \citep{1995ARA&A..33..581K}. If two host galaxies merge, as predicted by hierarchical galaxy formation models, the two black holes can form a supermassive black hole binary \citep{1980Natur.287..307B} via dynamical friction. Mergers of gas rich galaxies also drive a large quantity of gas in the centre, potentially forming massive, rotationally supported disc \citep{2005ApJ...630..152E,2007MNRAS.379..956D}. If this gas can cool efficiently, it settles into a thin accretion disc surrounding the black hole binary.

If the binary hardens up to sub-parsec scales through gravitational slingshot ejection of stars \citep{2003ApJ...596..860M} or other processes, such as the interactions with a gaseous accretion disc \citep{1999MNRAS.307...79I,2009MNRAS.398.1392L,2011MNRAS.412.1591N}, eventually gravitational waves (GW;  \citealt{PhysRev.136.B1224}) will carry away the remaining orbital energy and induce the coalescence in less than a Hubble time. The detection of these GWs is expected by proposed experiments such as the \emph{Laser Interferometer Space Antenna (LISA)}.

General relativity predicts \citep{2005PhRvL..95l1101P,2006PhRvD..74d1501C,2008PhRvD..78h1501T} that the gravitational waves emitted during the merger of the black holes carry away energy and momentum, thus resulting in a \emph{mass loss} and in a \emph{recoil} of the remnant black hole. Many authors \citep[e.g.][]{2008ApJ...684..835S,2010MNRAS.401.2021R,2010MNRAS.404..947C,2010A&A...523A...8Z} have explored the second possibility, with the final aim of predicting an electromagnetic (EM) afterglow of the coalescence. The detection of such an afterglow would help in constraining the properties of the merged black holes and of the host galaxies. In this paper we explore the first possibility, studying the response of the accretion disc to the mass loss. For the rest of this paper, we neglect the effect of recoil. Previous work  for this physical case have been done by \citet{2009PhRvD..80b4012M,2009ApJ...700..859O,2010MNRAS.404..947C} by means of numerical simulations. In the present paper, we present an analytical model for the surface density evolution of the disc following the merger, derived under the assumption of a collisionless disc. To assess the validity of our model, we compare it with the outcome of 3D numerical hydrodynamical simulations, using the Smoothed Particle Hydrodynamics (SPH) code \textsc{phantom} \citep{2010MNRAS.406.1659P,lodatoprice2010}.

The geometry of the accretion disc before the merger has been discussed by \citet{2002ApJ...567L...9A} and by \citet{2005ApJ...622L..93M}. We assume that the plane of the disc coincides with the orbital plane of the binary \citep{1997MNRAS.285..288L,1999MNRAS.307...79I}. For such a configuration, the secondary black hole will open a gap \citep{1994ApJ...421..651A} in the disc, creating a hollowed region of size approximately $2a$, where $a$ is the semi-major axis of the binary. The evolution of the system will then progress at the viscous timescale, the disc and the binary being in contact due to tidal interactions. When the gravitational wave emission sets in, the structure of the circumbinary disc and of the binary rapidly decouple, and the coalescence occurs on a very fast timescale compared to the disc dynamical one. We can then assume that at the moment immediately preceding the merger the disc is in rotation around a point mass, whose magnitude is given by the total mass of the binary. For our purposes, the merger of the two black holes can be described as an instantaneous reduction of the mass of the central object.

The paper is organised as follows. In Section 2, we derive an analytical model for the evolution of the surface density of the disc after the merger, and thus the mass loss, occurs. The model is then compared with numerical simulations, described in Section 3; we present our results in section 4. Finally, in Section 5, we draw our conclusions.

\section{Analytical model}

In this section we develop a 1D analytical model for the surface density evolution of the accretion disc in the case of mass loss. We make the assumption that the disc is composed of test particles, and we neglect the effects of pressure; thus we deal with a collisionless disc. This is the same approach that \citet{2008ApJ...676L...5L} followed for analysing the response to a black hole recoil (in that case through the use of numerical simulations). Since real discs are collisional, the validity of our model has to be investigated and will be discussed in section 4.

\subsection{Derivation}

Before the mass loss happens, we make the simplifying hypothesis that the test particles in the disc are on circular orbits. That is, we assume a Keplerian rotation curve with a rotation speed given by $V_\rmn{K}=\sqrt{GM/R}$, where $G$ is the gravitational constant, $M$ the total mass of the two merging black holes and $R$ the distance of the particle from the center of mass of the black holes in the disc plane. The specific angular momentum $j$ of each particle is given by $j=V_\rmn{K} R=\sqrt{GMR}$, or, conversely, the orbital radius as a function of the specific angular momentum reads
\begin{equation}
R=\frac{j^2}{GM}.
\label{eq:circul_radius}
\end{equation}

The merger of the two black holes, and thus the mass loss, happens at $t=0$, and causes a fractional change in the mass of the black holes of magnitude $\epsilon\ll 1$. Thus for $t>0$ the new mass of the remnant $M'$ is given by $M'=M(1-\epsilon)$.

The mass loss does not change the specific angular momentum of the particles, as happens in the case where a recoil is present \citep{2010MNRAS.401.2021R}. However, due to the change in mass, every particle initially at radius $R'$ will have a circularisation radius expressed by:
\begin{equation}
\rcirc(R')=\frac{j^2}{GM(1-\epsilon)}\approx R' (1+\epsilon).
\end{equation}

Thus, coherently with a \emph{decrease} in the central mass, the circularisation radius \emph{increases} by the same fractional amount, so that  after the merger the particles are in elliptical orbits. To first order in $\epsilon$, these orbits can be described using the epicycle approximation. This means that the radius oscillates in an harmonic fashion about the new circularisation radius with the epicyclic frequency $\kappa$. The distance of the particle from the center can be then expressed as a function of the initial radius $R'$ and of time as:
\begin{eqnarray}
R_{\rm{new}}(R',t)=\rcirc(R') + \nonumber \\
A (\rcirc(R')) \sin (\kappa t + 3 \pi /2),
\end{eqnarray}
where $A(\rcirc(R'))$ is the amplitude of the epicyclic oscillations. The initial phase is required to ensure that at $t=0$ the particle is at $R=R'$ and its radial velocity vanishes. These requirements also sets the amplitude of the oscillations, that is:
\begin{equation}
A=\rcirc-R'=\epsilon \rcirc.
\end{equation}

Finally, we recall that for a Keplerian disc the epicyclic frequency is equal to the orbital frequency $\Omega$. Evaluating this quantity at the circularisation radius one obtains:
\begin{equation}
\kappa=\Omega=\sqrt{\frac{GM(1-\epsilon)}{\rcirc^{3}}}.
\end{equation}

We have now all the ingredients to compute how the mass loss changes the surface density $\Sigma$ of the disc given the initial surface density $\Sigma_0(R)$. Under the constraint that the mass of test particles does not change, we can evaluate the required quantity as follows:
\begin{equation}
\Sigma(R,t)=\int_{R_{\rm{in}}}^{R_{\rm{out}}} \frac{R'}{R} \Sigma_0(R') \delta[R-R_{\rm{new}}(R',t)] \mathrm{d} R',
\label{eq:sigma_phys}
\end{equation}
where the limits of the integral are the inner $R_{\rm{in}}$ and the outer $R_{\rm{out}}$ radius of the disc, $\delta$ is the Dirac function and the extra $R'/R$ factor accounts for the surface element variation and enforces the conservation of mass. The Dirac function in the integral takes care of selecting the contributions to the surface density of particles, initially at radius $R'$, that at time $t$ are at radius $R$.

%With the purpose of simplifying the equations we are dealing with, we introduce now dimensionless units. We introduce a length scale $\xi$, that we leave as a free parameter, to scale the length unit, so that $r=R/\xi$. We choose units in which $G=1$, so that the time scale reduces to the dynamical time scale at $r=1$, i.e. $t_m=[\xi^3 / (G M)]^{1/2}$. The free length parameter can be regarded also as a free velocity parameter, that is, the keplerian velocity at $r=1$.
%
%With this units equation \eqref{eq:sigma_phys} becomes:
%\begin{equation}
% \Sigma(\hat{r},\hat{t})=\int_{\hat{r}_{in}}^{\hat{r}_{out}} \frac{\hat{r}'}{\hat{r}} \Sigma_0 (\hat{r}') \delta \left[ \hat{r} - \hat{r}' (1+\epsilon) - \epsilon \hat{r}' \sin \left(\frac{(1-\epsilon)^{0.5}}{(\hat{r}' (1+\epsilon))^{1.5}} t + 3 \pi / 2\right) \right] \mathrm{d} \hat{r}',
%\end{equation}

\subsection{Implied timescale}
\label{sec:timescale}
There is no analytic solution for the inversion of the argument of the Dirac delta function. However, we show that our model predicts the formation of density peaks in the disc. This is a consequence of the strong radial dependence of the epicyclic frequency with radius, $\kappa\propto R^{-3/2}$. This implies that, given two particles at a certain distance $\Delta R$, they will quickly go out of phase. When the phase difference between two given particles is of order $\pi$, one particle will be \emph{increasing} its radius while the other one is \emph{diminishing} it. If they are close enough, they will intersect, forming a region of high density. Conservation of mass implies that we expect also regions of depression.

With a simple argument we can derive a timescale $t_{\rm{dp}}$ for the formation of these density peaks. The formation of a peak occurs when the phase difference $\Delta \phi$ is of the order $\pi$ for particles that are $2 \epsilon \rcirc \simeq 2 \epsilon R$ apart, that is, the maximum initial distance for which they can intersect. Then we can write:
\begin{equation}
\Delta \phi = t_\mathrm{dp} [\kappa(R)-\kappa(R+2 \epsilon R)] =\frac{3 t_\mathrm{dp}  \epsilon}{t_{\rm dyn}},
\end{equation}
where $t_\mathrm{dyn}=1/\Omega$ is the dynamical timescale. Solving for $t_\mathrm{dp}$ we obtain, neglecting factors of order unity:
\begin{equation}
t_\mathrm{dp}\simeq \frac{t_\mathrm{dyn}}{\epsilon},
\label{eq:peak_development}
\end{equation}
This is roughly the same time scale found by \citet{2008ApJ...684..835S}; however we stress that in this section we built a time-dependent model that gives the time evolution of the density and not only the timescale for the formation of density peaks.

To get an estimate for this timescale, we can use the \citet{2002ApJ...567L...9A} model for the decoupling separation of the binary $a$, that corresponds roughly to the inner radius of the disc. Their expression is however valid only for the case of extreme mass ratio binary. Using their arguments, one can derive the following expression, that has a general validity:
\begin{equation}
a=\left( \frac{32}{15\sqrt{2}} \right)^{2/5} \alpha^{-2/5} \left(\frac{H}{R}\right)^{-4/5} q^{2/5} (1+q)^{-4/5} \frac{2 G M}{c^2} .
\label{eq:a_inner}
\end{equation}
Here $\alpha$ is the \citet{shakurasunyaev73} parameter for viscosity, $H/R$ is the aspect ratio of the disc prior to decoupling and $q$ is the mass ratio of the binary: $q=M_2/M_1$, where $M_1$ and $M_2$ are the masses of the two black holes.

For fiducial parameters of $\alpha=10^{-2}$, $H/R=10^{-2}$, $M=10^6 M_\odot$ and $q=1/2$, we obtain an inner disc radius of $\simeq 2 \ \mathrm{AU}$, so that we expect $t_\mathrm{dp} \simeq 3 \ \mathrm{days}$ at the inner disc radius for $\epsilon=0.05$. Note that in units of $2 GM/c^2$, that is approximately the Schwarzschild radius $R_\mathrm{s}$ of the remnant, $a \simeq 100 R_\mathrm{s}$ for a large range of values of $q$, due to its weak dependence on $q$.

\begin{figure}
\begin{center}
\includegraphics[width=\columnwidth,clip=true]{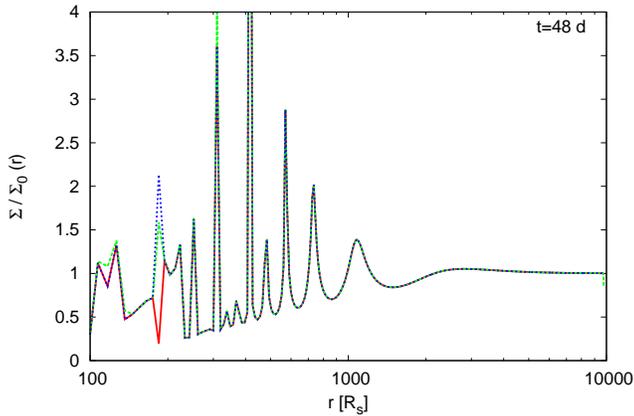}
\end{center}
\caption{Numerical solution, normalized to the initial unperturbed value, at time $t=48 d$ after the merger (red solid line). Sharp over-density peaks develop in the disc, as well as region of depression. The figure also shows a resolution study changing the $\tau$ parameter that controls the Gaussian width $\sigma$: dashed line green is the result for $\tau=10^{-4}$ and blue dotted line for $\tau=10^{-3}$, while the red line is the result for our fiducial value for $\tau=10^{-5}$. Outside 200 $R_\mathrm{s}$ we find excellent convergence, while some difference remains at the smallest radii.}
\label{fig:analytical_comp_res}
\end{figure}

\subsection{Numerical solution}

Since the integral \eqref{eq:sigma_phys} has no analytic solutions, we computed it numerically. We describe here briefly the method we used. To deal with the Dirac delta function, we replaced it using a Gaussian, so that the integral becomes:
\begin{eqnarray}
\Sigma(R,t)=\frac{1}{\sqrt{2 \pi} \sigma}\int_{R_{\rm{in}}}^{R_{\rm{out}}} \frac{R'}{R} \Sigma_0(R') \times \nonumber \\
\exp \left[-\frac{(R-R_{\rm{new}}(R',t))^2}{2\sigma^2}\right] \mathrm{d} R',
\end{eqnarray}
where $\sigma$ is the Gaussian width. Since the actual integral is recovered in the limit in which $\sigma \to 0$, $\sigma$ is chosen to be much smaller than the radius $R$ at which we are interested in computing the surface density: $\sigma= \tau R$, where $\tau$ is an arbitrary small number (we used $10^{-5}$ in our computations). Once $\sigma$ has been chosen, the integral can be computed with standard numerical techniques (we used Simpson's rule). To ensure the discretisation does not significantly modify the final result, we used a sufficiently large number of sampling points such that the separation between them is much less than $\sigma$ (we used a $10^{-2}$ ratio in our computations). %The separation betweens the discrete integration points is then chosen accordingly to $\sigma$, in order to have separations much smaller than $\sigma$. %To speed up the computation, we truncate the support of the Gaussian function after an interval of $5 \ \sigma$.

Figure \ref{fig:analytical_comp_res} shows the numerical solution, normalized to the initial unperturbed value, at time $t=48 d$ after the merger (red solid line), for fiducial parameters of $M=10^6 M_\odot$ and $\epsilon=0.05$. The unperturbed surface density profile is taken to be $\Sigma(R) \propto R^{-p}$. As shown in \ref{sec:timescale}, sharp over-density peaks develop in the disc, as well as region of depression. The figure also shows a resolution study changing the $\tau$ parameter that controls the Gaussian width $\sigma$: dashed line green is the result for $\tau=10^{-4}$ and blue dotted line for $\tau=10^{-3}$. Outside 200 $R_\mathrm{s}$ we find excellent convergence, while some difference remains at the smallest radii.

\section{Simulation details}
In order to test the validity of the model developed in the previous section, we set up numerical hydrodynamic simulations. Simulations are capable of following the full 3D, possibly non-linear hydrodynamical evolution of the disc, and can then be used to quantify the limits of a collisionless model.

\begin{figure*}
\begin{center}
\includegraphics[width=.8\textwidth,clip=true]{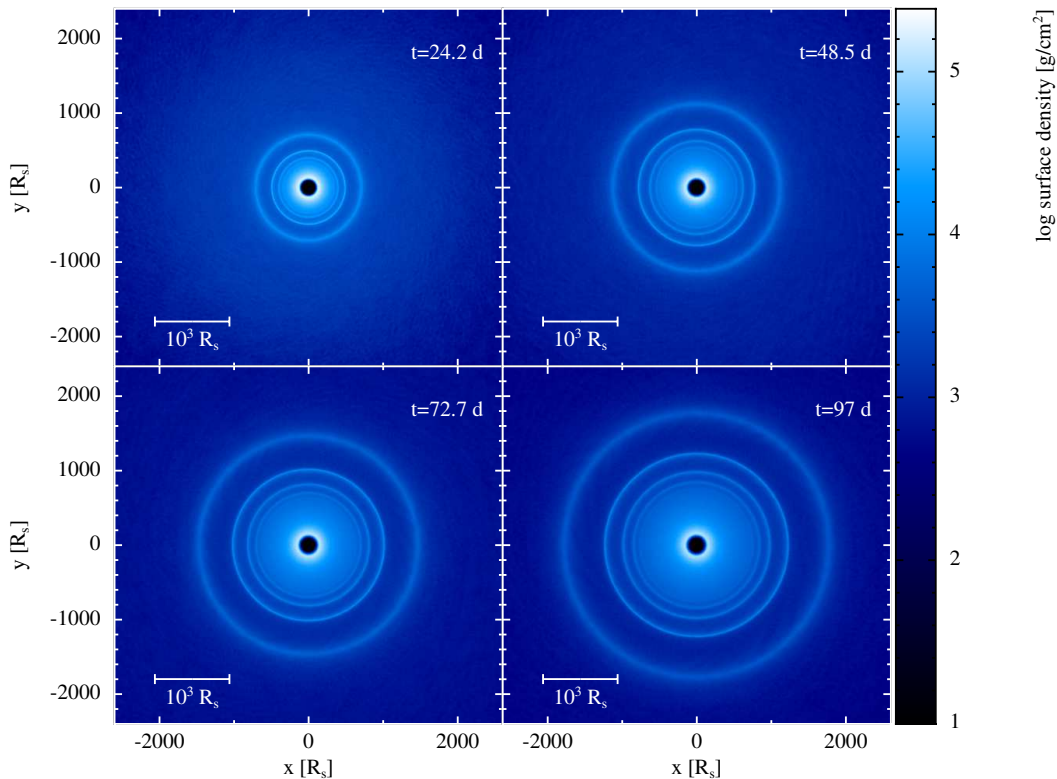}
\end{center}
\caption{Face-on view of the disc for simulation cold-iso at four different times. The rendered quantity is surface density, normalized to a total disc mass of $500 \ M_\odot$. The formation of sharp density peaks is particularly evident.}
\label{fig:iso_cold_facerender}
\end{figure*}

\begin{figure*}
\begin{center}
\includegraphics[width=.8\textwidth,clip=true]{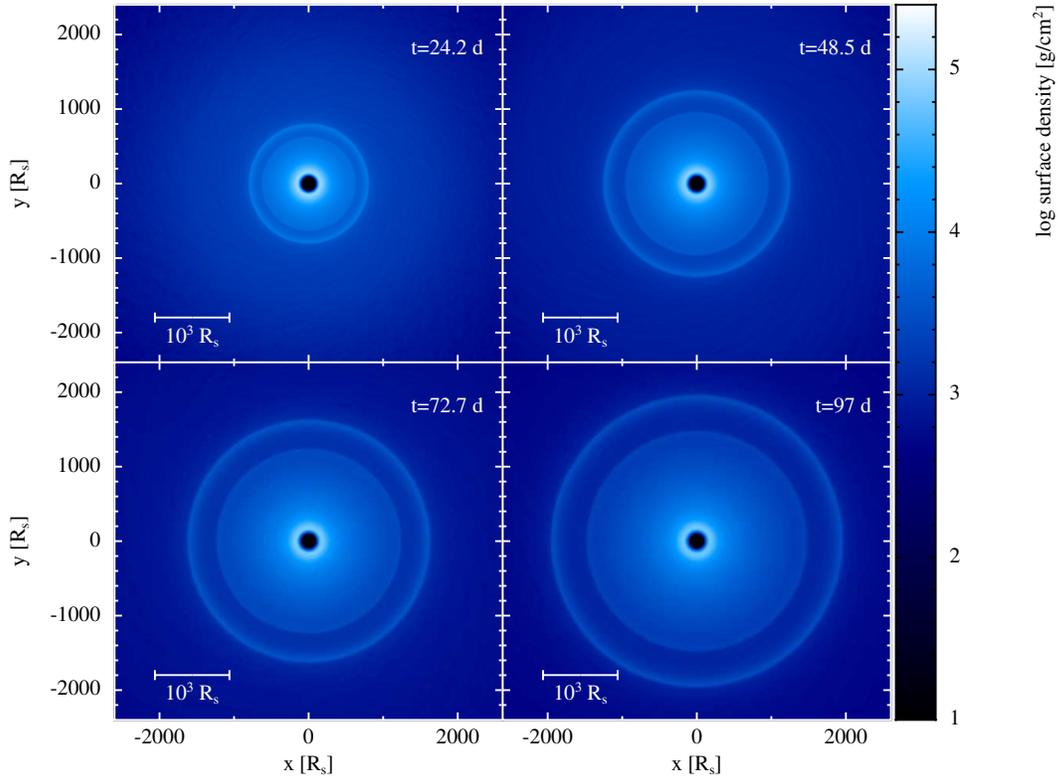}
\end{center}
\caption{Face-on view of the disc for simulation hot-iso at four different times. The rendered quantity is surface density, normalized to a total disc mass of $500 \ M_\odot$.}
\label{fig:iso_hot_facerender}
\end{figure*}

\begin{table*}
\centering
\begin{tabular}{cccccccc}
\toprule
Abbreviation & $R_\mathrm{out}$ & No. of particles & H/R & EOS & R/h  & H/h & $\alpha_\mathrm{SS,max}$\\
\midrule
cold-iso & 		$10^4$ & 2M &  	0.032 & 	Isothermal & 	$\sim 60$ & 2 & 0.025\\
cold-iso-hres &	$10^3$ & 2M &	0.032 &	Isothermal &	$\sim 300$ & 10 & 0.005\\
hot-iso &		$10^4$ & 7M & 	0.125 & 	Isothermal & 	$\sim 60$ & 8 & 0.00625\\
hot-iso-hres &	$10^3$ & 7M & 	0.125 & 	Isothermal & 	$\sim 300$ & 37 & 0.00125\\
cold-adiabatic &		$10^4$ & 2M & 	0.032 & 	Adiabatic & 	$\sim 60$ & 2 & 0.025\\
hot-adiabatic &		$10^4$ & 7M & 	0.125 & 	Adiabatic & 	$\sim 60$ & 8 & 0.00625\\
\bottomrule
\end{tabular}
\caption{The table summarizes the simulations run and the abbreviation used in the text to refer to them. We explored different temperatures of the disc and different equations of state. To check for numerical convergence, we increased radial resolution of our simulations by truncating the disc at the radius $R_\mathrm{out}$ indicated (in units of the Schwarzschild radius). The aspect ratio H/R and the radial resolution R/h are evaluated indicated in the table are evaluated at $10^2 R_\mathrm{s}$, while the vertical resolution H/h is indipendent of radius. Note that the radial resolution is an increasing function of radius, so that the value reported in the table is a lower limit for this value. For details about the equation of state, see section \ref{sec:ic}; about the radial resolution, section \ref{sec:res_study}. We report in the table also the values of $\alpha_\mathrm{SS,max}$, the maximum equivalent Shakura-Sunyaev parameter that is attained at shocks. The typical values of the effective $\alpha_\mathrm{SS}$ away from shocks are one or two orders of magnitude smaller.}
\label{tab:sim_run}
\end{table*}

\subsection{\textsc{phantom}}
Our simulation use the \textsc{phantom} SPH code. 
SPH \citep{1977MNRAS.181..375G,1990nmns.work..269B,monaghan05,PriceReview} is a well-known Lagrangian technique to solve the hydrodynamics equations. It is based on a discretisation of the fluid mass in terms of particles, where each particle represents a smeared out distribution of density. The formulation of SPH implemented in \textsc{phantom} is the so-called `grad-h' formulation \citep{2004MNRAS.348..139P,2007MNRAS.374.1347P}, which is based on a self-consistent derivation of the equations of motion from a least action principle. This guarantees an exact conservation of momentum, angular momentum, and energy. In this formulation the smoothing length $h$, which sets the effective resolution length of the simulation, and the density $\rho$ are mutually dependent, so that the smoothing length is adaptively adjusted to give a better resolution in the high density regions.

 \textsc{phantom} \citep{2010MNRAS.406.1659P,lodatoprice2010} is a low-memory, highly efficient 3D SPH code specifically designed for non self-gravitating problems.The code implements a linked list \citep{1985A&A...149..135M} for neighbour finding. The code has been parallelised using both OpenMP and MPI, although in this paper only the shared memory OpenMP parallelisation has been used.

\subsection{Initial conditions}
\label{sec:ic}
Our initial conditions consist of a thin disc \citep{1981ARA&A..19..137P}. Due to the arguments discussed in section 2.2, we set the inner radius of the disc to $100 \rschwarz$, where $R_\mathrm{s}$ is the Schwarzschild radius of the black hole. We set the outer radius to be $10^4 \rschwarz$.

The radial distributions of surface density and sound speed, $c_\mathrm{s}$, are power laws:
\begin{equation}
\Sigma(R) \propto R^{-p},
\end{equation}
\begin{equation}
c_\mathrm{s} (R) \propto R^{-q},
\label{eq:sound_speed}
\end{equation}
where the exponents $p$ and $q$ are chosen to satisfy the relation $p+2q=3$. This ensures that the vertical averaged smoothing length $\left<h\right>\propto H$, where $H$ is the height of the disc. This ensures that the disc is equally vertically resolved at each radius. For all our simulations we assume $p=3/2$ and therefore $q=3/4$. We will discuss in section 4 the normalisation of the sound speed chosen. The disc is isothermal in the vertical direction. Initial conditions were generated using random placement of particles.

We assume a Keplerian rotation curve, taking into account also the pressure gradient corrections, and we neglect relativistic corrections. A simple estimate using the well-known \citet{1980A&A....88...23P} potential gives that the error is of the order of a percent for the innermost radius of the disc, and even smaller for the rest of the disc. Thus the use of a Newton potential for modelling the black hole is sufficient.

For the thermodynamics, we experimented both with the use of an isothermal equation of state, where every particle keeps fixed its initial temperature given by equation \eqref{eq:sound_speed}, and the use of an adiabatic equation of state, where the internal energy of each particle can change. In both cases we assumed an ideal gas law. The two equations of state chosen correspond to the physical case of a very fast cooling timescale and a very long one. We choose the value $\gamma = 5/3$, corresponding to ideal monatomic gas. 

At time $t=0$, we assume that the central source of the gravitational field experiences a mass loss of fractional change $\epsilon=0.05$, and we let the system evolve in time. In section 2.2 we have shown that the timescales we are interested in are of the order of days, thus much shorter than the viscous timescales. For this reason we do not include physical viscosity in our simulations, and we include only the artificial viscosity needed to capture shocks in SPH. \textsc{phantom} includes a slightly modified version of the \citet{Monaghan1997} formulation of artificial viscosity, based on an analogy with Riemann solvers. Details on the implementation can be found in \citet{2010MNRAS.406.1659P}. We chose standard values to parametrize the viscosity, with $\alpha=1$ and $\beta=2$. The meaning of $\alpha$ can be understood by comparison with the compressible Navier-Stokes equation. It can be shown that in the case of constant $\alpha$ the artificial viscosity (e.g. \citealp{lodatoprice2010}) is equivalent to a Shakura-Sunyaev parameter $\alpha_\mathrm{SS}$, given by:
\begin{equation}
\alpha_\mathrm{SS}=\frac{1}{20} \alpha \frac{h}{H}.
\end{equation}
The corresponding bulk viscosity coefficient is a factor 5/3 greater
than the shear.

In our simulation we use the \citet{morris&monaghan1997} switch to reduce dissipation away from shocks, so that the individual $\alpha$ of each particle can vary between $0.01$ and $1$, so that the viscosity coefficients vary. In a presence of a shock front, however, the values of $\alpha$ are near the maximum, so that one can use the maximum value to compute the viscosity coefficients.

We report in table \ref{tab:sim_run} the values of $\alpha_\mathrm{SS,max}$ for the various
simulations, where $\alpha_\mathrm{SS,max}$ is the maximum equivalent
Shakura-Sunyaev parameter that is attained at shocks, where $\alpha=1$.
Since in our simulations we do not have strong shocks, the typical
values of the effective $\alpha_\mathrm{SS}$ are one or two orders of
magnitude smaller.
%\begin{equation}
%P=\frac{k_{\rm B}}{\mu m_{\rm H}} \rho T,
%\end{equation}
%where $k_{\rm B}$ is the Boltzman constant, $\mu$ the mean molecular weight, and $m_{\rm H}$ the hydrogen mass.

\begin{figure*}
\begin{tabular}{cc}
\includegraphics[width=.45\textwidth]{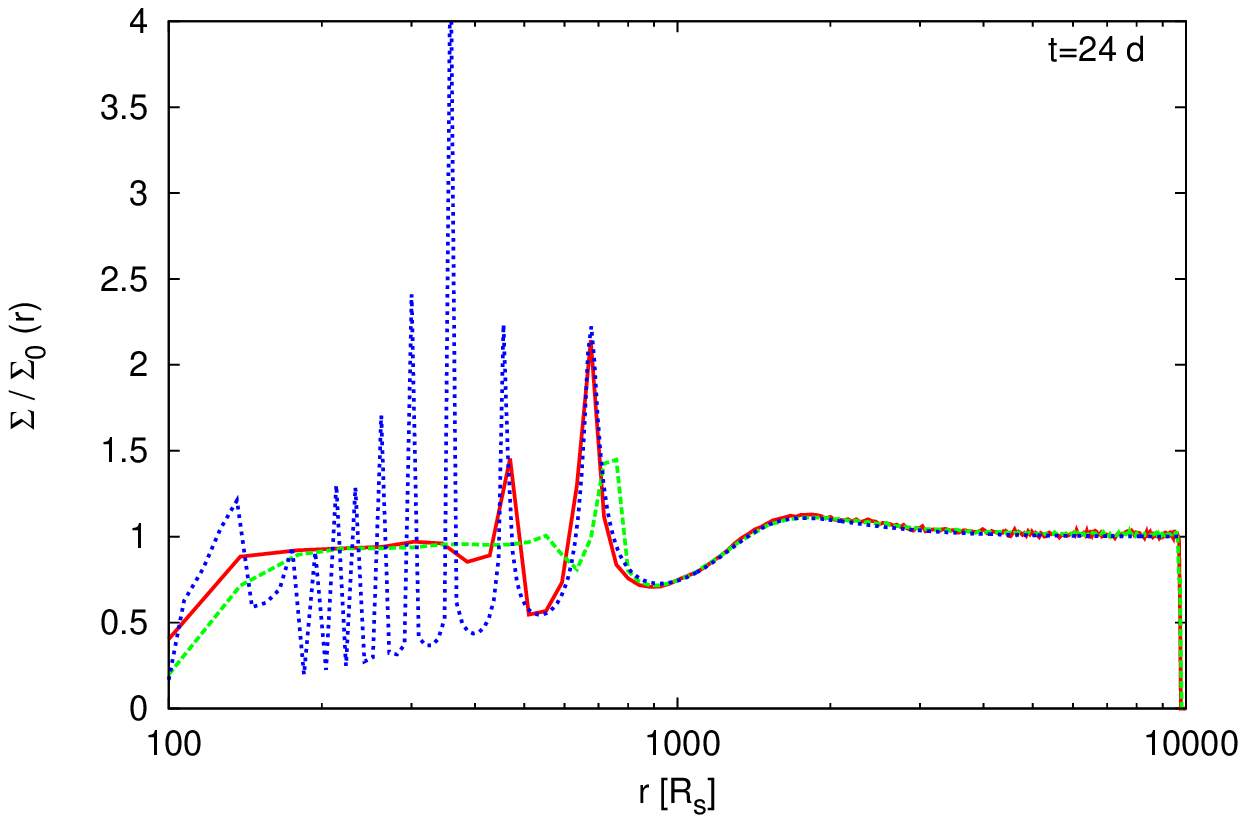}&
\includegraphics[width=.45\textwidth]{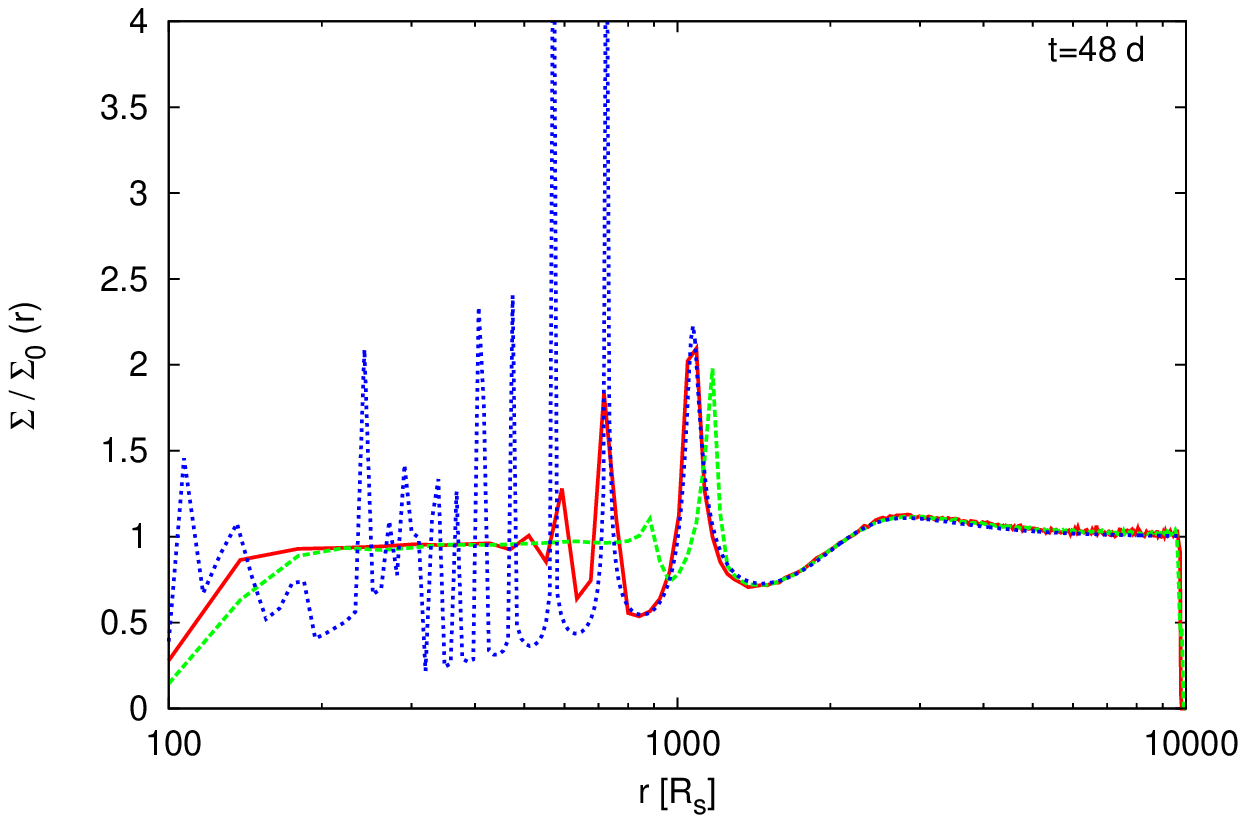}\\
\includegraphics[width=.45\textwidth]{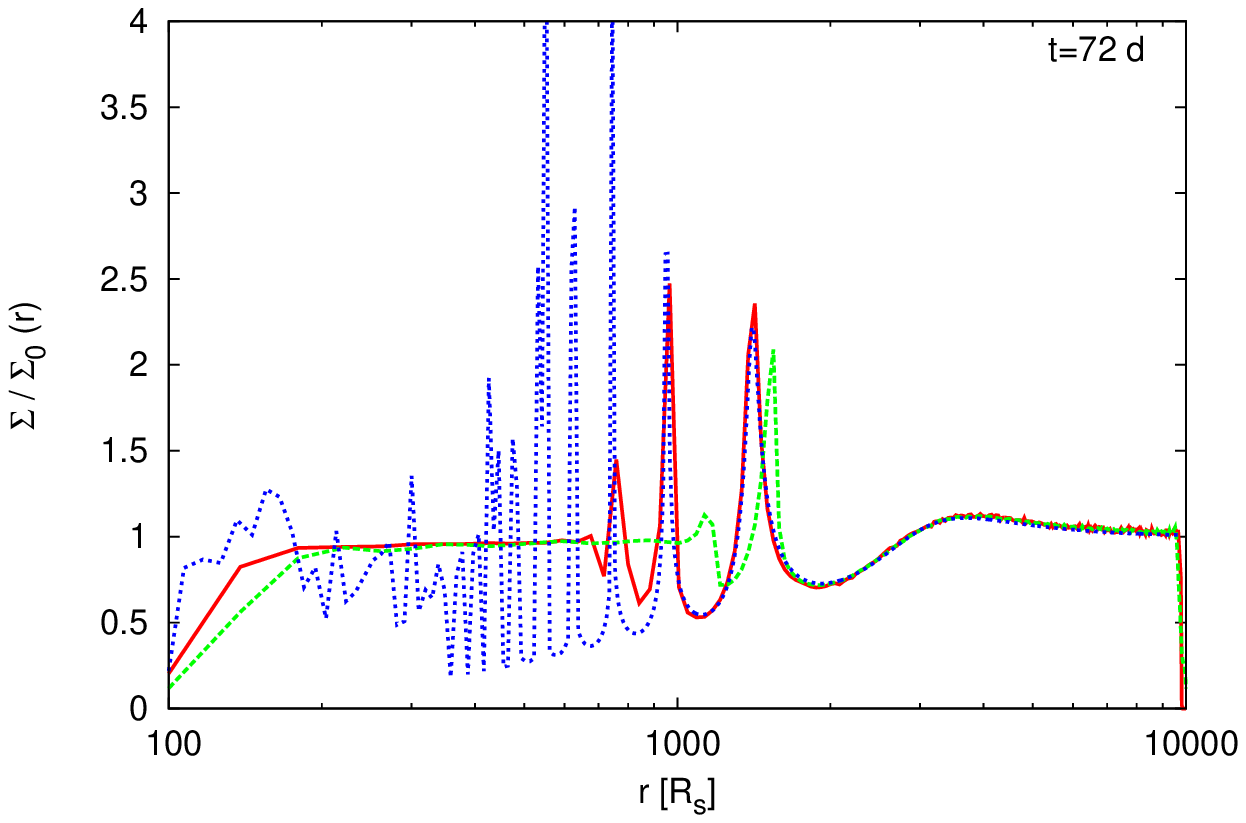}&
\includegraphics[width=.45\textwidth]{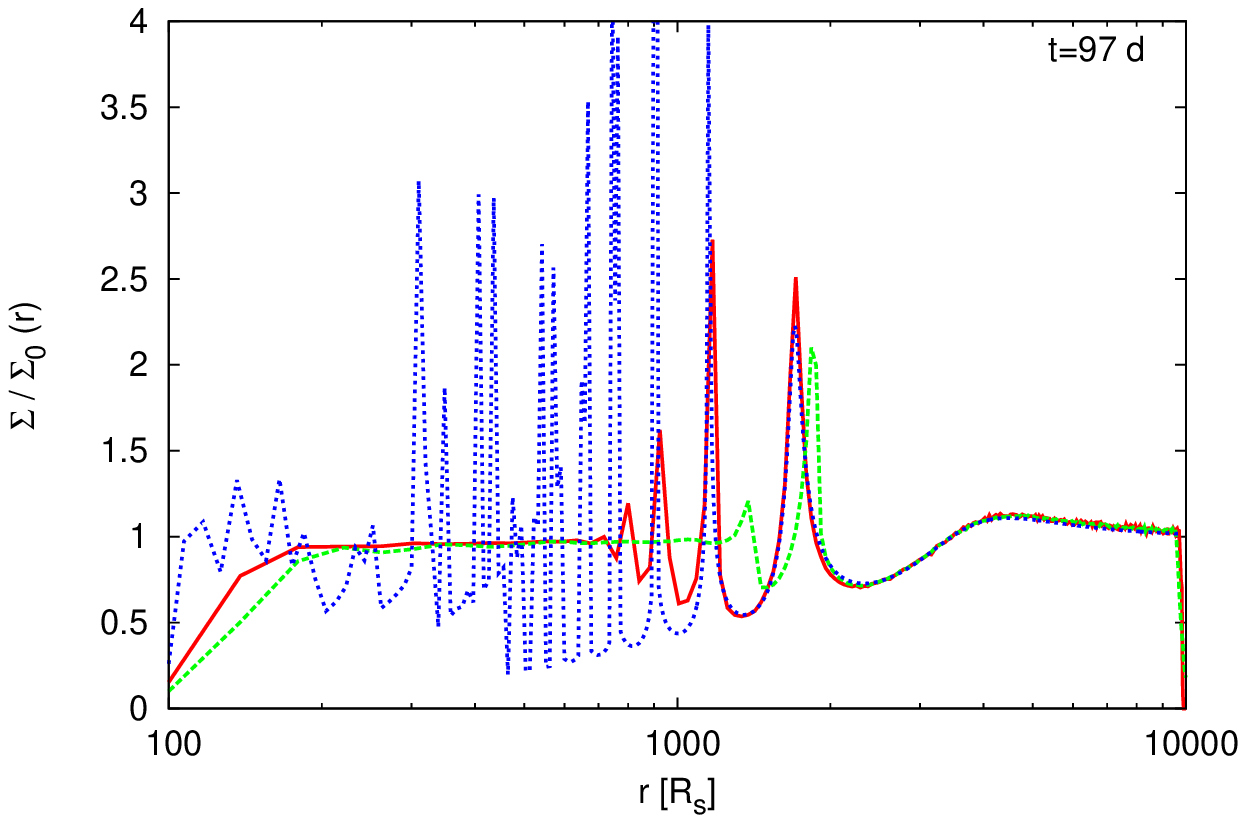}\\
\end{tabular}
\caption{Surface density normalized to initial value at four different times for simulations cold-iso (solid red line) and hot-iso (dashed green line), compared with the model (dotted blue line).}
\label{fig:iso_surfcomp}
\end{figure*}

\section{Results}

In this section we present our results and we discuss the comparison between our model and the simulations. We run simulations with four different sets of physical parameters, studying the dependence on the normalisation of the sound speed and on the equation of state. For the sound speed, we took two different values, that correspond roughly to an aspect ratio $H/R$ of the disc of $0.032$ and $0.125$ at $R=10^{2} R_\mathrm{s}$. In the rest of the disc, the aspect ratio exhibits a mild dependence on radius ($\propto R^{-1/4}$), given the chosen sound speed profile. We will refer in the rest of the paper to the two cases as ``cold'' and ``hot'' disc. 

We studied also the resolution dependence in our simulations to check that numerical convergence has been reached. To run simulations at a significantly higher resolution than our ``fiducial'' one, we simulate only the inner part of the disc, truncating it at $10^3 R_\mathrm{s}$, keeping the same number of particles. This ensures that the mass of each SPH particle becomes significantly lower. We summarize the simulations run in table \ref{tab:sim_run}.

Since we employed Schwarzschild units for the lengths, our results do not depend on the mass of the central black hole. To show physically meaningful times, however, in the following section time and temperature are expressed in physical units for the fiducial case in which $M=10^6 M_\odot$.

\subsection{Isothermal simulations}
\label{sec:iso}

We first present results from the isothermal simulations, namely cold-iso and hot-iso. Figure \ref{fig:iso_cold_facerender} and Figure \ref{fig:iso_hot_facerender} present face-on views of the disc, rendering surface density, at four different times for these two simulations. More quantitatively, Figure \ref{fig:iso_surfcomp} shows a comparison between the two simulations (cold-iso: solid red line; hot-iso: dashed green line) and the analytic model (dotted blue line) at four different times, namely t=24d, 48d, 72 and 97 d after the mass loss. We show surface density, normalized to the pre-massloss value, as a function of radius. 

The main effect visible is the formation of density peaks after the mass loss. Once formed, these peaks then propagate outwards. This is consistent with the prediction of the model that the peaks first form in the inner part of the disc, where the timescale for their formation is faster (see equation \ref{eq:peak_development}), and then travel outwards. There is a quite remarkable agreement between the model and the simulations in the outer part of the disc. We should stress that, since our analytic model is collisionless, the propagation of these peaks is not a hydrodynamical effect (like the propagation of waves in a fluid medium). Instead, this effect comes only from the dynamics of the particles comprising the disc. However, fluid effects in a real disc can modify the evolution in time, and this is the reason why we need also hydrodynamical simulations in addition to the analytical model. Before proceeding with a more detailed analysis, we note that also \citet{2010MNRAS.404..947C} run hydrodynamics simulations in the case of mass-loss, with a very different numerical scheme (they use the grid-code FLASH). Qualitatively, the outcome of their simulations (see their Figure 5) is very similar to what we have obtained. This a confirmation that the result does not depend on the numerical method employed.

We find that a fluid disc differs from the model in two main aspects:
\begin{enumerate}
\item the position of the peaks for the hot disc does not coincide with the one predicted by the model (see figure \ref{fig:iso_surfcomp});
\item the disturbances are damped in the fluid disc, so that in the inner part of the disc the analytical model and the simulations disagree. This effect is stronger for the hot disc.
\end{enumerate}

To account for the first difference, we can notice, using equation \eqref{eq:peak_development}, that disturbances travel at a speed $v_\mathrm{d}$ given by: 
\begin{equation}
v_\mathrm{d}=f \epsilon V_\mathrm{K},
\label{eq:peak_speed}
\end{equation}
 where $f$ is a number of order unity. On the other hand, no information in a fluid can travel faster than the sound speed $c_\mathrm{s}$. It is then relevant to consider the ratio $\mathcal{M}_\epsilon=\epsilon V_\mathrm{K} / c_\mathrm{s}$. If this ratio is greater than 1, the disturbance travels faster than the sound speed. In this regime, the inner fluid cannot inform the outer one of the disturbance, and the position of the peaks is the one given by the analytical model. In the opposite regime, the fluid will communicate to the outer disc the formation of the peak, thus distorting its shape and shifting it to outer radii.

\begin{figure}
\includegraphics[width=\columnwidth]{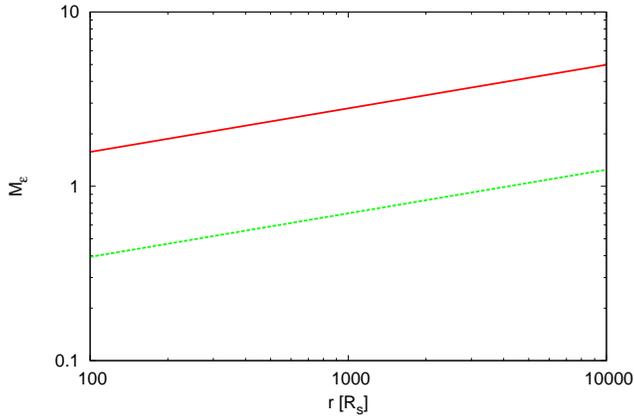}
\caption{The $\mathcal{M}_\epsilon$ parameter as a function of radius for the cold-iso (red solid line) and hot-iso (green dotted line) simulations. While for the cold disc the parameter is everywhere greater than 1, thus being in the regime in which the disturbance travels at high supersonic speed, in the hot case the parameter is slightly less than 1, and the disturbance travels at a speed comparable with the speed of sound.}
\label{fig:m_epsilon}
\end{figure}

In our discs, $\mathcal{M}_\epsilon \propto R^{-1/2}/R^{-3/4} \propto R^{1/4}$, with a value at $10^3 R_\mathrm{s}$ given by $2.8$ for the cold case and $0.7$ for the hot case. Figure \ref{fig:m_epsilon} shows $\mathcal{M}_\epsilon$ as function of radius for the two simulations. It can be readily seen that for the cold disc $\mathcal{M}_\epsilon$ is always greater than 1, and this is consistent with what we have pictured above. For the hot disc, the results are intermediate between the two extremes, and this accounts for the shift of the peaks and for the distortion of the shape.

\begin{figure}
\includegraphics[width=\columnwidth]{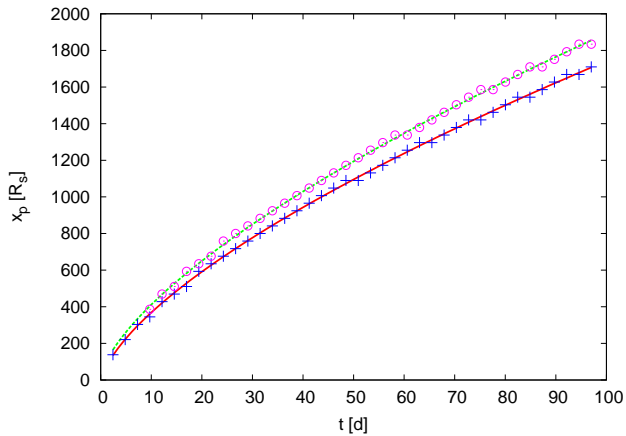}
\caption{Position of the outermost peak as a function of time for the cold-iso simulation (crosses) and for hot-iso (circles). Overplotted are the best fit to the points using equation \eqref{eq:r_t}. In the cold case we get $f=1.66$, while in the hot case we get a higher value of $f=1.87$. The value of $R_0$ we get from the fit is compatible in both cases, given the fit uncertainties, with zero.}
\label{fig:peak_position}
\end{figure}

In particular, if the velocity of the peak is given by equation \eqref{eq:peak_speed}, the position of the peak as a function of time is given by
\begin{equation}
R_\mathrm{p}(t)=\left(\frac{3}{2} \epsilon f t + R_0^{3/2}\right)^{2/3}.
\label{eq:r_t}
\end{equation}
Figure \ref{fig:peak_position} shows the position of the outermost peak as a function of time for the two simulations cold-iso (crosses) and hot-iso (circles). Overplotted are the best fit to the points using equation \eqref{eq:r_t}, where we let $f$ and $R_0$ be free parameters of the fit. In both cases there is a good agreement between the fit and the position of the peak. The value of $f$ we get from the fit are $1.66$ for cold-iso and $1.87$ for hot-iso, that is, in the hot disc case the peaks moves at a slightly higher velocity. These values also confirm that the constant $f$ is of order unity, and thus that $t_\mathrm{dp}$, from which equation \eqref{eq:peak_speed} has been derived, is a valid timescale for the formation of these density peaks. The value of $R_0$ we get from the fit is compatible in both cases, given the fit uncertainties, with zero.

Concerning the second difference, this is due to the fact that in the fluid disc the gas compression that occurs in the density peaks causes the energy of the epicyclic oscillations to be dissipated. Thus, for a given radius only a certain number of peaks can form before the oscillations are damped. This is due to a combination of shock dissipation and to $p \, \mathrm{d}V$ work. While in the isothermal case this work is lost from the system, in the adiabatic case it is converted into heat. As the gas temperature increase, the pressure term becomes more and more important, so that in the adiabatic simulations the oscillations are damped much faster than in the isothermal case. The pressure term also explains why in the hot simulations the perturbations are damped faster than in the cold one. Since the dynamical timescales are much faster in the inner disc, in the inner region the epicyclic motions already dissipated their energy, while in the outer region we can still see their effect.

\subsection{Resolution dependence}
\label{sec:res_study}

A very important issue is establishing if our simulations have reached numerical convergence. Too low resolution would imply that we cannot resolve properly some of the density peaks, thus being unable to distinguish between physical and numerical effects. We now turn to this problem.

A helpful parameter to consider is the radial resolution $R/h$, where $h$ is the SPH smoothing length. This is connected to the vertical resolution through the $H/R$ aspect ratio, so that the radial resolution can be expressed as $R/h= (H/R)^{-1} H/h$. It is then not possible, for a given aspect ratio, to have arbitrary values of the radial and vertical resolution. In this paper, when comparing simulations with different aspect ratios, we used the same radial resolution, so that the numerical uncertainties are the same for the two cases. Thus, the differences seen in the previous section between the hot and cold cases cannot be due to a resolution problem. Table \ref{tab:sim_run} also reports the radial resolutions of our simulations. Notice that since the aspect ratio is a function of radius, the radial resolution depends on radius (while our setup ensures that $H/h$ = const). However, the scaling is very mild and does not affect this discussion. In addition, we checked that all of our simulations are vertically resolved, with $H/h$ at least 2.

\begin{figure}
%\begin{tabular}{cc}
\includegraphics[width=.45\textwidth]{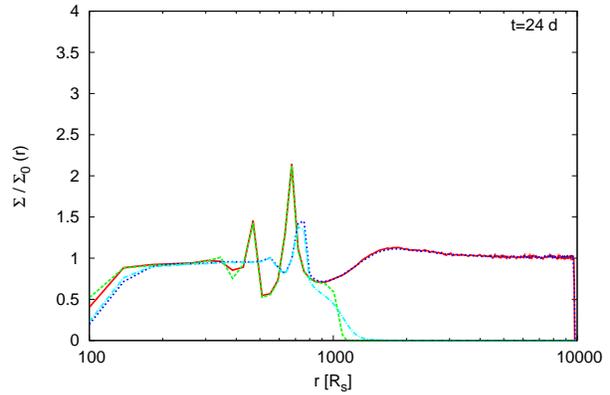}%&
%\includegraphics[width=.45\textwidth]{fig4b.eps}\\
%\end{tabular}
\caption{Surface density normalized to initial value at time $t=24 \mathrm{d}$ for simulations cold-iso (solid red line), cold-iso-hres (dashed green line), hot-iso (blue dotted line) and hot-iso-hres (light blue dotted-dashed line). No significant difference is found increasing the resolution.}
\label{fig:isor_surfcomp}
\end{figure}

\begin{figure*}
\begin{tabular}{cc}
\includegraphics[width=.45\textwidth]{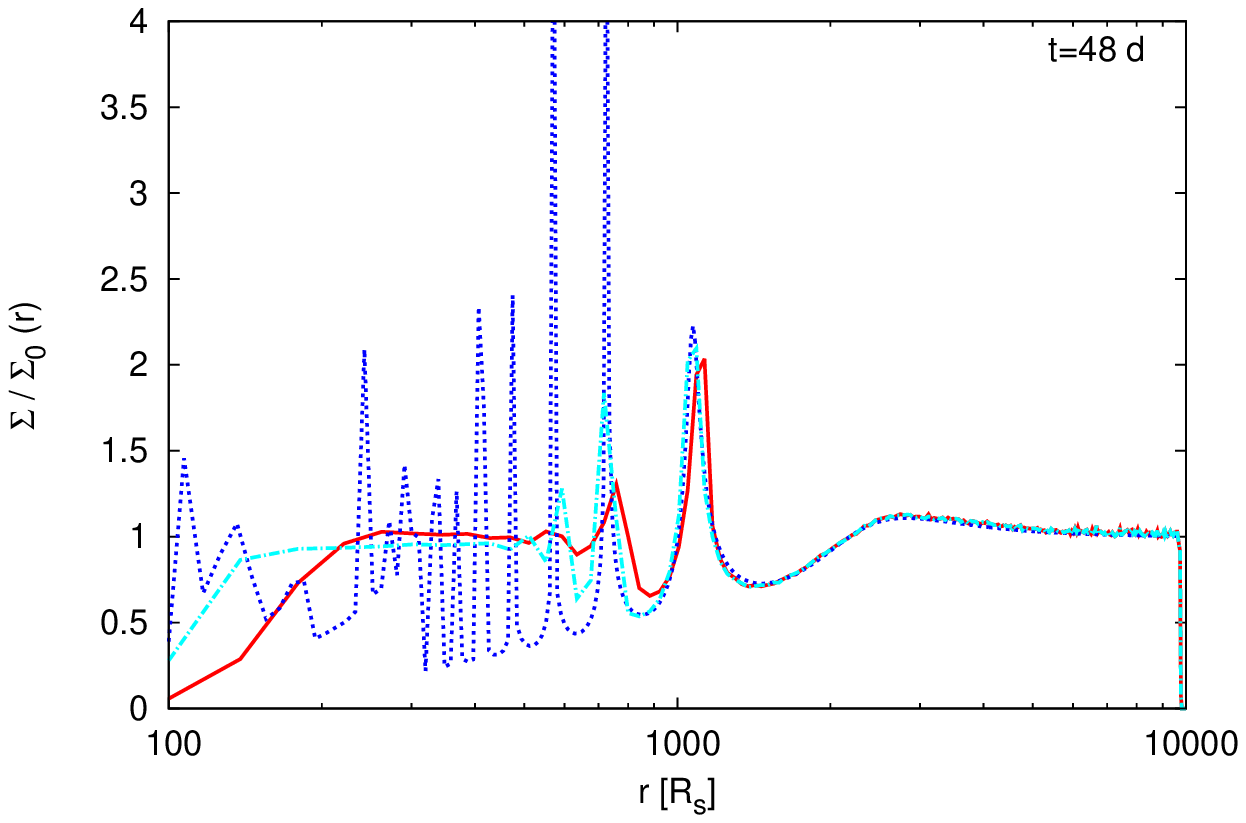}&
\includegraphics[width=.45\textwidth]{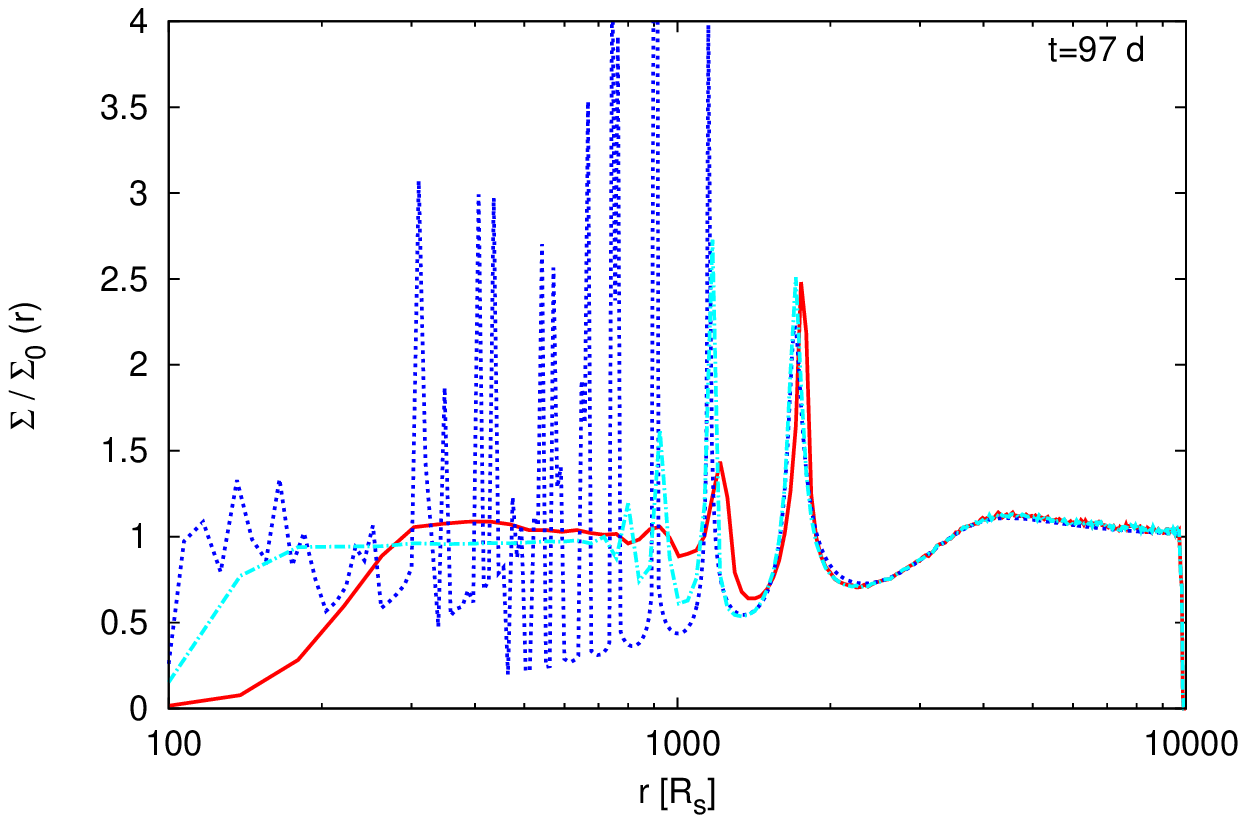}\\
\end{tabular}
\caption{Surface density normalized to initial value at two different times for simulations cold-adiabatic (red solid line), cold-iso (light blue dotted-dashed line) and for the analytical model (blue dotted line). Comparing the adiabatic with the isothermal case, both the position and the shape of the peaks are changed, as well as their height.}
\label{fig:adc_surfcomp}
\end{figure*}

\begin{figure*}
\begin{tabular}{cc}
\includegraphics[width=.45\textwidth]{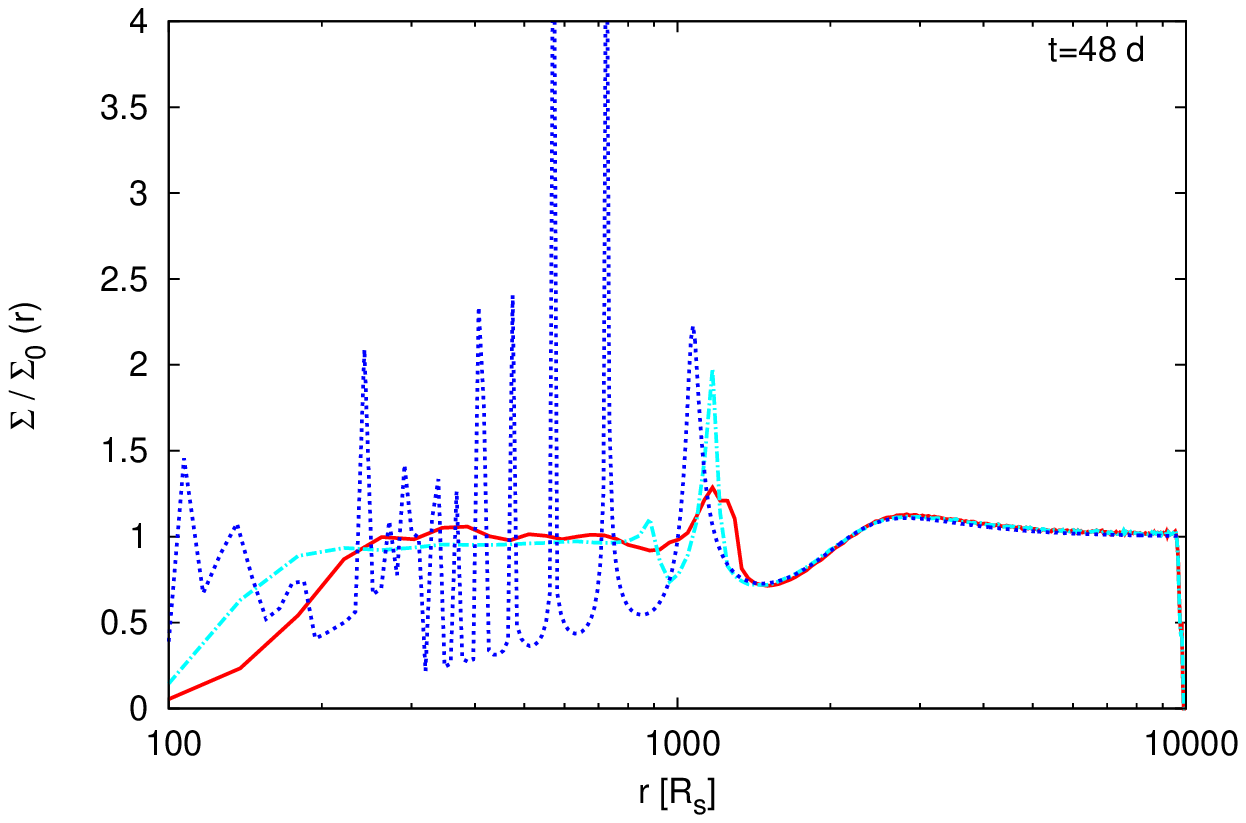}&
\includegraphics[width=.45\textwidth]{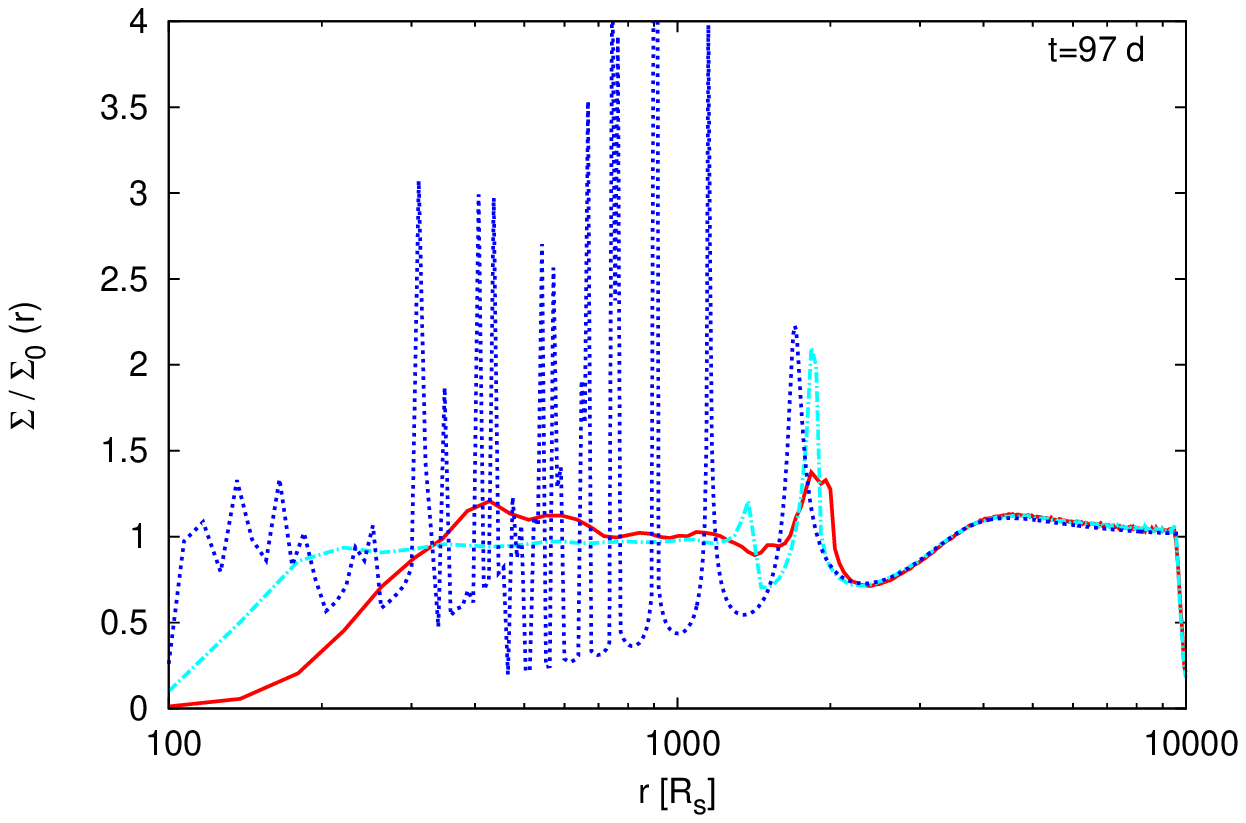}\\
\end{tabular}
\caption{Surface density normalized to initial value at two different times for simulations hot-adiabatic (red solid line), hot-iso (light blue dotted-dashed line) and for the analytical model (blue dotted line). Comparing the adiabatic with the isothermal case, both the position and the shape of the peaks are changed, as well as their height.}
\label{fig:adh_surfcomp}
\end{figure*}

To ensure that the radial resolution is high enough in our simulations, we resimulated the inner region of the disc employing the same number of particles used for our ``fiducial'' run. Figure \ref{fig:isor_surfcomp} presents a comparison between these high resolution runs, namely cold-iso-hres and hot-iso-hres, and the two simulations at standard resolution, namely cold-iso and hot-iso. We plot surface density as a function of radius at time $t=24 \mathrm{d}$. After this time, the outermost peak moves further than $1000 R_\mathrm{s}$, thus reaching the outer edge of the disc. We do not find significant differences between the two runs, even if there is a factor 5 of difference in radial resolution. We can then conclude that the effects described in the previous section are not due to a lack of resolution.

\subsection{Adiabatic simulations}

In order to study the dependence of the evolution of the disc on the thermal physics, we also run simulations using an adiabatic equation of state. In this case the gas can heat, and we expect thus a potential effect in the way the disturbances travel in the disc.

\begin{figure}
\includegraphics[width=\columnwidth]{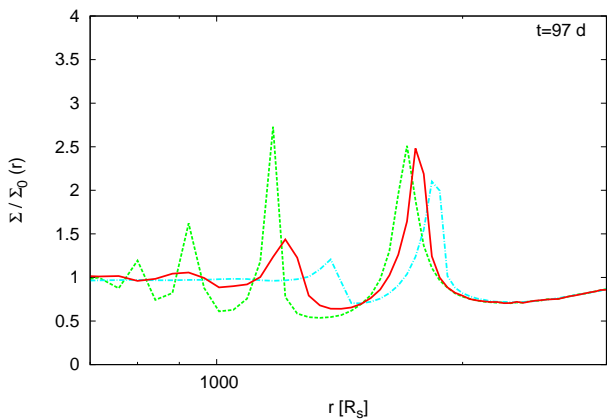}
\caption{Surface density normalized to initial value at two different times for simulations cold-iso (green dotted line), cold-adiabatic (red solid line) and hot-iso (light blue dotted-dashed line). The simulation cold-adiabatic shows a shift and a height of the peaks that is someway in the middle between the other two simulations.}
\label{fig:adzoom}
\end{figure}

In Figure \ref{fig:adc_surfcomp} we plot the surface density, normalized to the initial value, as a function of radius at time $t=48 \mathrm{d}$ and $t=97 \mathrm{d}$ for the simulations cold-adiabatic (red solid line), cold-iso (light blue dotted-dashed line), and for the analytical model (blue dotted line), while Figure \ref{fig:adh_surfcomp} plots the same quantity for simulations hot-adiabatic (red solid line), hot-iso (light blue dotted-dashed line) and for the analytical model (blue dotted line). It can be noticed that the both position and the shape of the peaks are changed with respect to the isothermal case, and also their height. In particular, the simulation cold-adiabatic shows a shift in the position of the peaks that is someway in the middle between the simulations cold-iso and hot-iso. To show this, we plot in Figure \ref{fig:adzoom} the results for simulations cold-iso (green dotted line), cold-adiabatic (red solid line) and hot-iso (light blue dotted-dashed line), zooming on the location of the peaks. It can be noticed that also the height of the peaks is smaller compared to the analytical model, and again intermediate between the simulations cold-iso and hot-iso.

\begin{figure}
\includegraphics[width=\columnwidth]{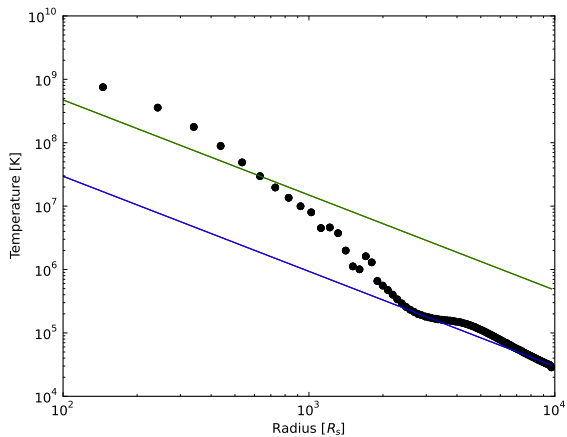}
\caption{Temperature as a function of radius for simulations cold-adiabatic (circles, computed by averaging the temperatures of individual SPH particles), hot-iso (green solid line) and cold-iso (blue solid line) at time $t=97 \mathrm{d}$. The temperature of the cold-adiabatic simulation at the position of the outermost peak is in the middle between the other two cases, thus accounting for the observed differences in the surface density evolution.}
\label{fig:cad_temperature}
\end{figure}

In order to account for these differences, we inspected the temperature structure of the cold-adiabatic simulation. In figure \ref{fig:cad_temperature} we plot the temperature of the disc as a function of radius (circles) at time $t=97 \mathrm{d}$, computed averaging the temperature of each SPH particle. For reference we also included the initial temperatures of the discs for the cold case (blue solid line) and for the hot one (green solid line). As can be seen, the disc heats up due to the shocks induced by the epicyclic motion. The total heating is not enough to reach the temperature of the hot case at the position of the outermost peak. Thus the $\mathcal{M}_\epsilon$ parameter introduced in section \ref{sec:iso} is intermediate between the values it assumes for the cold and the hot cases. This is consistent with the shape and position of the peaks observed, confirming that $\mathcal{M}_\epsilon$ is a good parameter to characterise the behaviour of the disc after the mass-loss. In the same spirit we can interpret also the behaviour of the hot-adiabatic simulation, that shows height, shape and positions of the peaks with stronger differences compared to the analytical model, since it can heat to even higher temperatures.

\section{Discussion and conclusions}

In this paper, we have investigated the evolution of the dynamics of an accretion disc after the massloss of two merging black holes. Under the assumption of a disc composed of test particles, we were able to derive an analytical model for the surface density evolution of the disc following the mass loss. The model predicts the formation of sharp density peaks in the disc. Once formed, these peaks travel outwards. However, we showed that they are not a type of hydrodynamical wave, but rather follow only from the kinematics of the disc. We derived also a timescale for the formation of these peaks, finding it of the same order of magnitude as previous findings. However, our model is fully time-dependent. 

To test the validity of our model, we set up numerical simulations, capable of taking into account the full hydrodynamics of the gaseous disc, using the SPH code PHANTOM. We found a good agreement in the shape and position of the peaks between the model and the simulations. There are however small differences, depending on the disc parameters, in the position and shape of the peaks from the model predictions. To account for these discrepancies, we introduced the $\mathcal{M}_\epsilon$ parameter. We showed that lower values of this parameter correspond to discs that deviate more strongly from the analytical model, because hydrodynamical effects are more important. We also showed that the timescale introduced is effective in predicting the formation of the density peaks. 

In the fluid disc, after some of the density peaks have formed, the epicyclic oscillations dissipate, thus causing the inner disc, where the dynamical timescales are faster, to significantly differ from the analytical model. We showed that the $\mathcal{M}_\epsilon$ parameter is also effective in explaining how fast this dissipation proceeds, that is, how many peaks form in the disc.

In our work we did not consider the radiation emitted by the fluid. We modelled two limiting cases, using an isothermal and an adiabatic equation of state, accounting for a very fast cooling timescale and a very long one, respectively. Further work is needed to quantify in which physical regime real discs lie. 

\section*{Acknowledgments}
Our visualisations made use of the SPLASH software package \citep{2007PASA...24..159P}. We thank Phil Armitage for interesting discussions and the referee for useful suggestions.

\bibliography{Rosottietal12}{}
\bibliographystyle{mn2e}

%\bibitem[\protect\citeauthoryear{Baird}{1981}]{b1} Baird S.R., 1981,
%ApJ, 245, 208
%\end{thebibliography}

\bsp

\label{lastpage}

\end{document}